\documentstyle[12pt]{article}
\begin{document}
\def\beq{\begin{equation}}
\def\eeq{\end{equation}}
\def\bea{\begin{eqnarray}}
\def\eea{\end{eqnarray}}
\def\ve{\vert}
\def\nnb{\nonumber}
\def\ga{\left(}
\def\dr{\right)}
\def\aga{\left\{}
\def\adr{\right\}}
\def\rar{\rightarrow}
\def\nnb{\nonumber}
\def\la{\langle}
\def\ra{\rangle}
\def\ba{\begin{array}}
\def\ea{\end{array}}

\title{ {\small { \bf THE STRONG COUPLING CONSTANTS OF EXCITED POSITIVE PARITY 
                      HEAVY MESONS IN LIGHT CONE  $QCD$} } }

\author{ {\small T. M. ALIEV \thanks {Permanent address: Institute of
Physics, Azerbaijanian Academy of Sciences, Baku, Azerbaijan}\,\,, N. K. PAK and 
M. SAVCI} \\ {\small Physics Department, Middle East Technical
University}\\ {\small 06531 Ankara, Turkey} }

\begin{titlepage}
\maketitle
\thispagestyle{empty}

\begin{abstract}
\baselineskip  0.7cm
We calculate the strong coupling constants $g_{P^{**}P^{*}\pi}$, where 
$P^{**}~(D^{**}, B^{**})$ is the $1^+$ p-wave state, in the framework of the
light cone $QCD$ sum rules, and using these  values of $g_{P^{**}P^{*}\pi}$,
we compute the hadronic decay widths for $D^{**} \rar D^*~\pi$ and 
$B^{**} \rar B^*~\pi$.

\end{abstract}
\end{titlepage}
\baselineskip  .7cm
\newpage

\setcounter{page}{1}
\section{Introduction}
The main goal of the future $\it c$ - $\tau$ and $B$ - meson factories
is a deeper and more comprehensive investigation of the properties of 
the heavy mesons, containing charm and beauty quarks. In particular, 
search for the excited states of $D$ and $B$ mesons and their decay 
modes constitute one of the main research program of the above 
mentioned factories, and can play an essential role for understanding 
the dynamics of the excited states.  

In general, for the interpretation of the experimental data from heavy 
meson physics, we need to know the large 
distance, i.e., nonperturbative effects. For example, for the 
exclusive decays which can easily be measured experimentally, we need 
a more accurate estimation of the form factors and other hadronic matrix 
elements that are described by the long distance effects,
thus one needs a method which 
takes into account the nonperturbative (long distance) effects. Among 
the different approaches in estimating the large distance effects, 
the $QCD$ sum rule method \cite{R1} occupies a special place, since 
this method is based on the  first principles of $QCD$ and 
the fundamental $QCD$ Lagrangian. 

In this work we use a version of the $QCD$ sum rule, namely the light cone
$QCD$ sum rule. This method is based on the {\it Wilson Expansion} of the 
$T$-product of currents near light cone, in terms of different non-local
operators. These operators are characterized by their twists rather than
their dimensions. Matrix elements of the non-local operators in the variable
external field are identified with a set of the wave functions of increasing
twist, and replace the vacuum expectation value of local operators that appear
in the traditional sum rule method. The form of wave functions are restricted 
by the conformal invariance of $QCD$. More about the details of this method
and its applications can be found elsewhere in the literature \cite{R2}-\cite{R19}. 

In this article this  method is used for the calculation
of the strong coupling constants $g_{P^{**}P^{*}\pi}$ of the excited positive parity 
meson decays, where $P^{**} (P^{*})$ is the $1^+ (1^-)$ meson state.
In sect.2 we derive the sum rule for the $g_{P^{**}P^{*}\pi}$ coupling constants.
Sect.3 is devoted to the numerical analysis, where we also compute the widths of
the $1^+$ meson decay to $P^*~\pi$ state.

\newpage

\section{Light Cone Sum Rule for $g_{P^{**}P^{*}\pi}$ Coupling Constant }
According to the general strategy of $QCD$ sum rule method, the coupling constant 
$g_{P^{**}P^{*}\pi}$ can be calculated by equating the representations of a suitable
correlator in hadronic and quark-gluon languages. For this aim we consider
the following correlator:
\beq
\Pi_{\mu\nu} = i \int d^4x~e^{ip_1x}~\la \pi(q) \ve T \aga \bar d(x) \gamma_{\mu} Q(x)
\bar Q(0) \gamma_{\nu} \gamma_5 u(0) \adr \ve 0 \ra~.
\eeq
Here $\bar d \gamma_{\mu} {\cal }Q (\bar Q \gamma_{\nu} \gamma_5 u )$
is the interpolating current for $1^-~(1^+)$ meson state, $Q$ is a heavy quark
(charm quark for $D$ meson, and beauty quark for $B$ meson case), $p_1$ is the 
momentum of the  $1^-$ meson.
When the pion is on the mass shell $q^2=m_{\pi}^2$, the correlator function (1) depends
on two invariants, $p^2$ and $p_1^2$. In what follows we set $m_{\pi}=0$.

First consider the physical (hadronic) representation of (1). Physical part of it 
can be expressed in terms of the contribution of the lowest lying resonances
$P^{**}$ and $P^{*}$ in the corresponding channels
\beq
\Pi_{\mu\nu} = \la \pi(q) P^*(p_1) \ve P^{**}(p) \ra~
\frac {\la P^* \ve \bar d \gamma_{\mu} Q \ve 0 \ra }{p^2_{1}-m_{P^*}^2}~
\frac {\la 0 \ve \bar Q \gamma_{\nu} \gamma_5 u \ve P^{**} \ra }{p^2-m_{P^{**}}^2}~.
\eeq
The matrix elements entering in eq.(2) are defined in the standard manner:
\bea
\la P^* \ve \bar d \gamma_{\mu} Q \ve 0 \ra &=& 
m_{P^*} f_{P^*} \epsilon_{\mu}(p_1)~, \nnb \\
\la 0 \ve \bar  Q \gamma_{\nu} \gamma_5 u \ve P^{**} \ra &=& m_{P^{**}} f_{P^{**}}
\epsilon_{\nu}^{(1)}(p)~,
\eea
where $m_{P^*}~(m_{P^{**}})$, $f_{P^*}~(f_{P^{**}})$ and $\epsilon_{\mu}~(
\epsilon_{\nu}^{(1)})$ are the mass, leptonic decay constant, and the polarization 
vector of the vector $1^-~(1^+)$ meson state.
In general, the matrix element $\la \pi P^* \ve P^{**}\ra$ can be written as
\beq
\la \pi(q) P^*(p_1) \ve P^{**}(p) \ra = F_0(\epsilon^* \epsilon^{(1)}) +
F_1(q \epsilon) (q \epsilon^{(1)})~.
\eeq
In obtaining eq.(4) we have used the transversality condition, $p_1 \epsilon = 
p \epsilon^{(1)} = 0$, and $p_1 = p-q$. The second term in eq.(4)  
gives negligible contribution to the decay width $P^{**} \rar P^* \pi$,
in comparision to the first term, since it is proportional to 
\bea
\ga m_{\pi}^2 + \frac { {\Delta}^4}{m_{P^{**}}^2 m_{P^*}^2} \dr ~, \nnb
\eea
\newpage

where $\Delta = m_{P^{**}}-m_{P^*}$ . Therefore we shall
neglect the second term in (4) and set $g_{P^{**}P^{*}\pi} \equiv F_0$.
Using eqs.(2), (3), and (4) for the physical part, we get 
\beq
\Pi_{\mu\nu} = \frac { m_{P^*} m_{P^{**}} f_{P^*} f_{P^{**}} g_{P^{**}P^{*}\pi }}
{(p^2-m_{P^{**}}^2) (p_1^2-m_{P^*}^2)} 
\aga g_{\mu\nu} - \frac { p_{\mu} p_{\nu} }
{m_{P^{**}}^2 } - \frac { p_{1\mu} p_{1\nu} }{ m_{P^*}^2} - 
\frac { (p_1p) p_{1\mu} p_{\nu} } { m_{P^*}^2 m_{P^{**}}^2 } \adr ~.
\eeq
At this point we would like to make the following comment. Since the vector current
$\bar q \gamma_{\mu} Q$ is not conserved, it also couples to  
$J^P=0^+$ scalar mesons $P_0$ as well as the $J^P=1^-$ vector mesons.
Therefore the $P_0$ contribution should be taken into account in the sum rule and 
this addition introduces further uncertainities. In order to avoid the $0^+$
meson contributions we must choose a structure that does not contain its 
contribution. 

Noting that the corresponding transition matrix element is given as 
\bea
\la 0 \ve \bar q \gamma_{\mu} Q \ve P_0 \ra = f_{D^0} m_{D^0} p_{1\mu}~,  \nnb
\eea
that is, only the structure $\sim$ $p_{1\mu}$ contains the $0^+$ meson 
contribution, it follows from eq.(5) that we have only two 
structures, namely $\sim$ $g_{\mu\nu}$ and $p_{\mu} p_{\nu}$, which do not 
contain the $J^P=0^+$ meson contribution.
In our analysis we choose the structure $\sim$ $g_{\mu\nu}$. We also perform
calculations for the structure $p_{\mu} p_{\nu}$ and the final
results of both structures in  predicting of $g_{P^{**} P^* \pi}$ are 
practically the same.

After performing double Borel transformation over variables $-p^2$ and $-p_1^2$ (see 
eq.(5)) for the physical part of the sum rule for $g_{P^{**}P\pi}$, for the 
structure of $g_{\mu\nu}$, we get :
\beq
\Pi^{\it phys}=
\frac{1}{M_1^2 M_2^2} g_{P^{**}P^*\pi} f_{P^{**}} f_{P^*} m_{P^{**}} m_{P^*} 
e^{\ga - \frac {m_{P^{**}}^2+m_{P^*}^2}{2 M^2} \dr}~.
\eeq
Now we turn out attention to the theoretical part of (1). In this calculation 
we will use the notation of the work \cite{R16}.
After a lengthy calculation
we get (after double Borel transformation over $-p^2$ and $-p_1^2$)
\bea
\Pi^{\it theor}&=&
f_{\pi} \frac{1}{M_1^2 M_2^2} e^{-\frac{m^2}{M^2}} 
\Bigg{[} \frac{m_{\pi}^2}{m_u+m_d} m_Q M^2 \varphi_P(u_0) \nnb \\
&&+~ 2 m^2_Q g_2(u_0) - \frac{1}{2} M^4 \varphi'_{\pi}(u_0) 
+2 (M^2+m^2_Q) (g'_1(u_0)+G'_2(u_0)) \nnb \\
&&+~\frac{1}{2} M^2 \Big{[}2\int_{0}^{u_0} d\alpha_1 
\int_{u_0-\alpha_1}^{1-\alpha_1} d\alpha_3 
\frac {\varphi_{\parallel}(\alpha_1,1-\alpha_1-\alpha_3,\alpha_3)}{\alpha_3^2} \nnb \\
&&+~ \int_{0}^{u_0} \frac {d\alpha_3}{\alpha_3}\aga
\tilde{\varphi}_{\parallel}(u_0-\alpha_3,1-u_0,\alpha_3) -
\varphi_{\parallel}(u_0-\alpha_3,1-u_0,\alpha_3)\adr \nnb \\ 
&&-~ \int_{0}^{1} \frac {d\alpha_3}{\alpha_3} \aga
\tilde{\varphi}_{\parallel}(u_0,1-u_0,\alpha_3) + \varphi_{\parallel}(u_0,1-u_0-\alpha_3,\alpha_3)
\adr \Big{]} \Bigg{]} \nnb\\ 
&& +~ \mbox{(continuum contribution)}~.
\eea
The pion wave functions $\varphi_\pi(u)$, $\varphi_P(u)$, $g_1(u)$ and $G_2(u)$,
have the twists $\tau = 2,~\tau=3$, $\tau=4$ and $\tau=4$ respectively, and they 
appear in the matrix elements of nonlocal quark operators as shown below (see \cite{R6}
 and \cite{R16}):
\bea
\la \pi(q) \ve \bar d \gamma_{\mu}\gamma_5 u(0) \ve 0 \ra &=&
-i f_{\pi} q_{\mu} \int_{0}^{1} du~e^{iqux}\Big{[} \varphi_{\pi}(u) +
x^2 g_1(u) + {\cal O}(x^4) \Big{]} \nnb \\ 
&&+ f_{\pi} \ga x_{\mu} - \frac{x^2 q_{\mu}}{q x}
\dr \int_{0}^{1} du~e^{iqux} g_2(u)~,
\eea

\beq
\la \pi(q) \ve \bar d i \gamma_5 u(0) \ve 0 \ra =
\frac{f_{\pi} m_{\pi}^2}{m_u+m_d} \int_{0}^{1} du~e^{iqux} \varphi_{P}(u)~,
\eeq 

\beq
G_2(u) = - \int_{0}^{u} g_2(v) dv~.
\eeq
The functions $\varphi_(\alpha_i)$ and $\tilde{\varphi}_(\alpha_i)$ are the
twist-4 wave functions and are defined in the following way:
\bea
&&
\la \pi(q) \ve \bar d(x) \gamma_{\mu}\gamma_5 g_s G_{\alpha \beta}(ux) u(0) 
\ve 0 \ra = \nnb \\
&& f_{\pi} \Bigg{[} q_{\beta} \ga g_{\alpha \mu} - \frac {x_{\alpha} q_{\mu}}
{qx} \dr - q_{\alpha} \ga g_{\beta \mu} \frac {x_{\beta} q_{\mu}}{qx} \dr
\Bigg{]} \int {\cal D}\alpha_i \varphi_{\perp}(\alpha_i) e^{iqx(\alpha_1+u \alpha_3)} \nnb \\
&& +~ f_{\pi} \frac{q_{\mu}}{qx} \ga q_{\alpha} x_{\beta} - q_{\beta} x_{\alpha} \dr
\int {\cal D}\alpha_i \varphi_{\parallel}(\alpha_i) e^{iqx(\alpha_1+u \alpha_3)}~,
\eea

\bea
&&
\la \pi(q) \ve \bar d(x) \gamma_{\mu}g_s \tilde{G}_{\alpha \beta}(ux) u(0)
\ve 0 \ra = \nnb \\
&& i f_{\pi} \Bigg{[} q_{\beta} \ga g_{\alpha \mu} - \frac {x_{\alpha} q_{\mu}} 
{qx} \dr - q_{\alpha} \ga g_{\beta \mu} \frac {x_{\beta} q_{\mu}}{qx} \dr
\Bigg{]} \int {\cal D}\alpha_i \tilde{\varphi}_{\perp}(\alpha_i) e^{iqx(\alpha_1+u \alpha_3)} \nnb \\
&&+~ i f_{\pi} \frac{q_{\mu}}{qx} \ga q_{\alpha} x_{\beta} - q_{\beta} x_{\alpha} \dr
\int {\cal D}\alpha_i \tilde{\varphi}_{\parallel}(\alpha_i) e^{iqx(\alpha_1+u \alpha_3)}~,
\eea
and 
\bea
\tilde{G}_{\alpha \beta} &=& \frac{1}{2} \epsilon_{\alpha \beta \sigma \lambda}
G^{\sigma \lambda}~,~~~\mbox{and}~~~ \nnb \\
{\cal D}\alpha_i &=& d\alpha_1 d\alpha_2 d\alpha_3 \delta(1-\alpha_1-\alpha_2
-\alpha_3)~. \nnb
\eea
In (7) we set
\bea
M^2=\frac{M_1^2 M_2^2}{M_1^2 + M_2^2}~ ,~~~ u_0 = \frac{M_1^2}{M_1^2 + M_2^2}~~~
\mbox{and}~~~ \varphi' = \frac{d\varphi}{du} {\ve}_{u=u_0}~. \nnb 
\eea

\newpage

We omitted the path-ordered factor $ P e^{ [ i g_s \int_{0}^{1} du~x^{\mu} 
A_{\mu}(ux) ]} $, since in the Fock-Schwinger gauge $x^{\mu} A_{\mu} = 0$,
it is trivial. In numerical calculations, we choose the symmetric point $u_0=\frac{1}{2}$ which means 
that quark and antiquark have equal momenta inside the pion. At this point the  
subtraction of the continuum can be done by substituting
\bea
e^{- \frac{m_Q^2}{M^2}} \rar e^{- \frac{m_Q^2}{M^2}} - 
e^{- \frac{s_0^2}{M^2}}~,
\eea
at least for the twist-3 contribution \cite{R16}. However, 
we use this substitution everywhere in (7) 
since higher twist contributions are negligible.
Equating eqs.(6) and (7) and using (13) 
we finally obtain the following sum rule for the strong coupling constant $g_{P^{**}P\pi}$ :
\bea
g_{P^{**}P\pi} f_{P^{**}} f_{P^*} &=&
\ga \frac{1}{m_{P^{**}} m_{P^*}} \dr e^{ \ga \frac {m_{P^{**}}^2+m_{P^*}^2}{2 M^2} \dr}
f_{\pi} M^2 \ga e^{- \frac{m_Q^2}{M^2}} -
e^{- \frac{s_0^2}{M^2}} \dr \times \nnb \\
&&\Bigg{[}\frac{m_{\pi}^2}{m_u+m_d} m_Q \varphi_P(u_0) + 
\frac {2 m_Q^2}{M^2} g_2(u_0) - \frac{1}{2} M^2 \varphi'_{\pi}(u_0) \nnb \\
&&+~ 2 \ga 1+ \frac{m_Q^2}{M^2} \dr (g'_1(u_0)+G'_2(u_0)) \nnb \\
&&+~ \frac{1}{2} \Big{[}2\int_{0}^{u_0} d\alpha_1 \int_{u_0-\alpha_1}^{1-\alpha_1} d\alpha_3
\frac {\varphi_{\parallel}(\alpha_1,1-\alpha_1-\alpha_3,\alpha_3)}{\alpha_3^2} \nnb \\
&&+~\int_{0}^{u_0} \frac {d\alpha_3}{\alpha_3}\aga\tilde{\varphi}_{\parallel}(u_0-\alpha_3,1-u_0,\alpha_3) -
\varphi_{\parallel}(u_0-\alpha_3,1-u_0,\alpha_3)\adr \nnb \\
&&-~ \int_{0}^{1} \frac {d\alpha_3}{\alpha_3} \aga
\tilde{\varphi}_{\parallel}(u_0,1-u_0,\alpha_3) + \varphi_{\parallel}(u_0,1-u_0-\alpha_3,\alpha_3)
\adr \Big{]} \Bigg{]}~.
\eea
From (14) it follows that, for the calculation of  
the value of the strong coupling constant $g_{P^{**}P^*\pi}$, we need to know 
the leptonic decay constants of the $P^{**}$ and $P^*$, $f_{P^{**}}$, and $f_{P^*}$.
The decay constant
$f_{P**}$ can be obtained from the two point sum rules:
\bea
f_{P^{**}}^2 m_{P^{**}}^2 &=& \frac {1}{8 \pi^2} \int_{m_Q^2}^{s_0} ds
e^{ \ga \frac {m_{P^{**}}^2 - s}{ M^2} \dr} \frac { (s-m_Q^2)^2}{s}
\ga 2 + \frac {m_Q^2}{s} \dr  \nnb \\
&& +~m_Q \la \bar q q \ra e^{ \ga \frac {m_{P^{**}}^2 - m_Q^2}{M^2} \dr}
\ga 1 - \frac {m_0^2 m_Q^2}{4 M^2} \dr~,
\eea
where
\bea
m_0^2= \frac{\la \bar q \sigma_{\alpha \beta} G^{\alpha \beta} q \ra }{\la \bar q q \ra } 
 = (0.8 \pm 0.2 )~~GeV^2~. \nnb
\eea
\newpage
To obtain $f_{P^*}$ it is necessary to make the
following replacements: $m_{P^{**}} \rar m_{P^*}$ and change the sign in front of the second term 
in (15) (in this case,  of course the value of the continuum threshold must also change ).  
Note that we do not take into account the perturbative ${\cal O}(\alpha_s)$ corrections 
in (15), as they  are not included in (14) either.
Note that the values of the decay constants $f_{B^*}$ and $f_{D^*}$ we use in our
calculations are given in \cite{R16}.

\section{Numerical Analysis}
For the numerical analysis of the $QCD$ sum rule (14) we first give the values of the input 
parameters:
\bea 
&&
f_{\pi}= 132~ MeV,~~ \frac {m_{\pi}^2}{m_u+m_d}(1~ GeV) = 1.65~GeV,~~ m_c=1.3~ GeV, \nnb \\
&&m_b=4.7~ GeV,~~ m_{D^{**}} = 2.420~ GeV,~~ m_{D^*}= 2.01~ GeV, \nnb \\
&&m_{B^*}=5.279~ GeV,~~ m_{B^{**}} = 5.732~ GeV, \nnb \\
&&(s_0)_D = 6 \div 8~ GeV^2,~~ (s_0)_B = 35 \div 40~GeV^2~. \nnb 
\eea
Using these parameters, from (15) for the leptonic decay constants $f_{D^{**}}$ and $f_{B^{**}}$
we get 
\bea
&&
f_{D^{**}}=(300 \pm 30)~MeV~, \\
&&
f_{B^{**}}=(200 \pm 20)~MeV~ .
\eea

Sum rule for $g_{P^{**}P^{*}\pi}$ contains the nonperturbative quantities, namely the wave 
functions. In our numerical analysis we use the wave functions proposed in \cite{R6}
(see also \cite{R16}). Having the values of the input parameters, one must find the region of 
Borel parameter
$M^2$, for which the sum rule eq.(14) is reliable.The lowest value of $M^2$ is usually fixed 
by imposing the condition that the terms proportional to the $\frac{1}{M^2}$ are resonably 
small. The upper bound for $M^2$ is usually fixed by the condition that the continuum 
and higher states contributions constitute about $(25 \div 30\%)$ of the ground resonance 
contribution. Under these conditions the fiducial range of $M^2$ for $B~(D)$ case
turns out to be $~8~GeV^2<M^2<20~GeV^2~(2~GeV^2<M^2<6~GeV^2)$. 
Using the values of the input parameters we get
\bea
&&
f_{B^{**}} f_{B^*} g_{B^{**}B^{*}\pi} = (0.78 \pm 0.12)~GeV^3~, \\
&&
f_{D^{**}} f_{D^*} g_{D^{**}D^{*}\pi} = (0.68 \pm 0.10)~GeV^3~.
\eea
\newpage
Dividing this product by the decay constants, we finally obtain for the 
$D^{**}D^*\pi~(B^{**}B^*\pi)$ coupling constants :
\bea
&&
g_{B^{**}B^{*}\pi} = 24 \pm 3~GeV~, \\
&&g_{D^{**}D^{*}\pi} = 10 \pm 2~GeV~.
\eea

Substituting these values in the expressions for the 
decay widths,

\bea
\Gamma(P^{**0} \rar P^{*+}~\pi^-) &=& \frac { g^2_{P^{**}P^{*}\pi} }{24\pi}
\ga 2 + \frac { (m^2_{P^{**}}+m^2_{P^*})^2 } 
{4 m^2_{ P^{**}} m^2_{P^*} } \dr \times \nnb \\
&&\frac {\Big{[} \aga m^2_{P^{**}} - (m_{P^*}+m_{\pi})^2 \adr 
\aga m^2_{P^{**}} - (m_{P^*}-m_{\pi})^2 \adr \Big{]}^{\frac{1}{2}} } 
{ 2 m^3_{P^{**}} } 
\eea
we get,
\bea
&&
\Gamma(D^{**0} \rar D^{*+}~\pi^-) \simeq 249~MeV~, \nnb \\
&&\Gamma(B^{**0} \rar B^{*+}~\pi^-) \simeq 296~MeV~.
\eea
Strong coupling constants (and correspondingly the decay widths) of the decays
$P^{**-} \rar P^{*0}~\pi^-$, $P^{**-} \rar P^{*-}~\pi^0$ and
$P^{**0} \rar P^{*0}~\pi^0$ can easily be obtained from 
$P^{**0} \rar P^{*+}~\pi^-$ with the help of the isotopic invariance.

\end{document}